\documentstyle[preprint,aps,prl]{revtex}
\begin{document}
\draft
\title{ Density-Functional Theory of the nonlinear optical susceptibility:
application to cubic semiconductors }
\author{Andrea Dal Corso}
\address{Institut Romand de Recherche Num\'erique en Physique des Mat\'eriaux
(IRRMA), IN Ecublens, 1015 Lausanne, Switzerland.}
\author{Francesco Mauri}
\address{Department of Physics, University of California at Berkeley,
Berkeley, CA 94720, USA \\
and Materials Science Division, Lawrence Berkeley Laboratory, Berkeley,
CA 94720, USA}
\author{Angel Rubio}
\address{Departamento F\'\i sica Te\'orica. Universidad de Valladolid.
E-47011 Valladolid. Spain.}
\maketitle

\begin{abstract}
We present a general scheme for the computation of the time dependent (TD)
quadratic susceptibility ($\chi^{(2)}$) of an extended insulator obtained
by applying the `$2n+1$' theorem to the action functional as defined
in TD density functional theory.
The resulting expression for $\chi^{(2)}$ includes self-consistent
local-field effects, and is a simple function of the linear response of the
system. We compute the static $\chi^{(2)}$ of nine III-V and
five II-VI semiconductors using the local density approximation(LDA)
obtaining good agreement with experiment.
For GaP we also evaluate the TD $\chi^{(2)}$ for second harmonic generation
using TD-LDA.
\end{abstract}

\pacs{42.65.Ky,71.10.+x,78.20.Wc}
\narrowtext

Nonlinear optics is a growing field of research which has applications
in many technical areas such as optoelectronics, laser science, optical
signal processing and optical computing~\cite{Physicstoday}.
In these fields the description of several physical phenomena, such as
optical rectification, wave-mixing, Kerr effect or multi-photons
absorbtion, relies on the knowledge of the nonlinear optical (NLO)
susceptibilities.
Moreover nonlinear spectroscopy is a powerful
tool to analyze the structural and electronic properties of extended and
low dimensional systems.
In the present work we give a general scheme to compute from first principles
the time dependent (TD) quadratic susceptibility ($\chi^{(2)}$)
of real materials within
TD-density functional theory (DFT).
Futhermore we show that the values of the static $\chi^{(2)}$ obtained
in the local density approximation (LDA) are in good agreement with
measured values for the cubic semiconductors.
Our approach makes feasible the computation of $\chi^{(2)}$
in cells containing up to an hundred atoms, since
it requires the same numerical effort as
the computation of the total energy.
This allows the evaluation of $\chi^{(2)}$ for systems of technological
and scientific relevance which can not be handled by the traditional methods,
such as surfaces or crystals of organic molecules.

Nowadays many first-principle calculations for the ground
state properties of materials are performed within
DFT.
Even in its simplest form, namely in the LDA
for the exchange and correlation energy this scheme gives results
which, in many cases, are in surprisingly good agreement with experiments.
A rigorous extension of DFT to TD phenomena
has been proposed in Ref.s~\cite{tddftperiodic,tddft}.
Although the available approximations for the exchange and correlation
energy are less accurate in the TD domain than in the static case,
this scheme is sufficiently general to allow many possible improvements
in the future.
Therefore TD-DFT seems to be a promising framework for the study of the
NLO susceptibilities.

Standard quantum-mechanical perturbation theory can be used to compute
the $\chi^{(2)}$.
The straightforward application
of perturbation theory leads to an expression for $\chi^{(2)}$,
which diverges for an infinite solid in the static limit.
However, for an insulator, these divergences have been shown to be
apparent~\cite{Sipe}.
This kind of approach has been applied to compute
the $\chi^{(2)}$ from first principles.
The non self-consistent expression for $\chi^{(2)}$ reported in
Ref.~\cite{Sipe} has been evaluated by
Huang and Ching~\cite{cinesi} using the DFT-LDA wavefunctions and
eigenvalues.
A fully self-consistent theory of the NLO
susceptibility within DFT has been proposed in a series of
papers by Levine and Allan~\cite{Levine}.
Their method is feasible but algebraically very involved
due to the necessity of dealing with the second order perturbation of the
wavefunctions and with the apparent divergences.
Their final expression is not easy to handle and its evaluation
requires summations over the conduction band states, which are
time consuming and difficult to converge.

In a previous paper two of us~\cite{Andrea} have shown that it is convenient
to regard the static $\chi^{(2)}$ as a third order derivative
of the total energy with respect to
an uniform electric field. We pointed out that this derivative can be obtained
by combining a Wannier representation of the electronic wavefunctions with the
`$2n+1$' theorem of perturbation theory~\cite{Langhoff,Gonze}.
We also found an equivalent expression of the static $\chi^{(2)}$
in terms of Bloch wavefunctions.

In the present letter we show that the method of Ref.~\cite{Andrea}
applies also to TD
periodic perturbations and to the self-consistent TD-DFT functional.
The TD $\chi^{(2)}$ can be regarded as a third order derivative of the
total action.
The stationary principle for the action
functional~\cite{tddftperiodic,tddft}, which
replaces in the TD case the miminum principle for
the energy functional, allows the use of the `$2n+1$' theorem.
As in the static case the third order derivative depends only on
the unperturbed wavefunctions and on their first order change due to the
TD electric field.
All the self-consistent contributions are included in the formalism in
a simple way. The final expression avoids perturbation sums and
does not present any apparent divergency.
We apply our formalism to the computation of the
the static $\chi^{(2)}$ of nine III-V and five II-VI
cubic semiconductors within the LDA.
For GaP we also evaluate the TD $\chi^{(2)}$ for second-harmonic
generation (SHG) using TD-LDA~\cite{zang}.

In the Kohn and Sham (KS) formulation of DFT
the ground state density $n^g({\bf r})$ of a system of $N$
interacting electrons in an external potential $V_{ext}({\bf r})$
is written in terms of $N/2$ single particle wave-functions $\{\phi^g\}$.
The set $\{\phi^g \}$ minimizes the KS energy functional $E[\{\phi\}]$
and the ground state energy is obtained as $E^g=E[\{\phi^g\}]$.
A formalism similar to that of the static case can be introduced also in the
TD domain if one restricts to Hamiltonians periodic in
time and to the evolution of the system which is steady and has the same
periodicity of the Hamiltonian~\cite{Sambe}.
In TD-DFT the TD steady density $n^s({\bf r},t)$
of a system of $N$ interacting
electrons in an external TD potential
$V_{ext}({\bf r},t)$, periodic in time with period $T$,
is expressed in terms of a $N/2$ TD
single particle wave-functions $\{\psi^s \}$~\cite{tddftperiodic,tddft}.
The set $\{\psi^s \}$ make stationary the KS action functional
$A[\{\psi \}]$, i.e.
\begin{equation}
\delta A[\{\psi^s \}]/ \delta \langle \psi_k(t)| = 0 ,\label{sc}
\end{equation}
and the steady action is obtained as $A^s =A[\{\psi^s \}]$.
The KS action functional $A[\{\psi \}]$ is defined as (atomic units are used
throughout) :
\begin{eqnarray}
A[\{\psi \}]   &=&
\int^T_0{dt\over T}
\left[ \sum_{i=1}^{N/2} 2\langle \psi_i(t)|-{1\over 2}\nabla^2
                        -i{\partial\over\partial t}|\psi_i(t)\rangle +
\int d^3r V_{ext}( { \bf r },t)n({\bf r},t)
                                                      \right]\nonumber \\
&&
+A_H[n]+A_{xc}[n]. \nonumber
\end{eqnarray}
Here $|\psi_i(t)\rangle =|\psi_i(t+T)\rangle$,
$\langle\psi_i(t)|\psi_j(t)\rangle = \delta_{ij}$,
$A_H[n]=\int^T_0dt/T\int d^3r d^3r'
n({\bf r},t)n({\bf r}',t)/( 2|{\bf r}-{\bf r}'| )
$ is the Hartree functional,
$A_{xc}[n]$ is the exchange and correlation functional, and
$ n({\bf r},t)= \sum_{i=1}^{N/2}2\langle
\psi_i(t)|{\bf r}\rangle \langle {\bf r}|\psi_i(t)\rangle $, where the
$2$ factor is for spin degeneracy.
At this stage no approximation for
the exchange and correlation functional is made.
The stationary principle in Eq.~(\ref{sc}) yields the TD KS
equations:
$$
i{\partial\over \partial t}|\psi_k^s(t)\rangle=[H_{KS}(t)-\epsilon_k]
|\psi_k^s(t)\rangle, \nonumber
$$
here $\epsilon_k$ are the steady states eigenvalues,
$H_{KS}(t)=-{1\over 2}\nabla^2 +V_{ext}({\bf r},t)+V_{Hxc}[n]({\bf r},t)$
is the time dependent KS Hamiltonian, and
$V_{Hxc}[n]({\bf r},t)=T\delta(A_H[n]+A_{xc}[n])/\delta n({\bf r},t)$.

Now we consider a potential of the form
$V_{ext}({\bf r},t,{\bf a})=V^0_{ext}({\bf r})+a_1 {\bf e}_1 \cdot {\bf r}\cos
(\omega_1t)
+a_2 {\bf e}_2 \cdot {\bf r}\cos (\omega_2t)+a_3 {\bf e}_3\cdot
{\bf r}\cos (\omega_3t)$,
where ${\bf e}_1$, ${\bf e}_2$, ${\bf e}_3$ are unit vectors
describing the orientation of three TD uniform electric fields,
$\omega_1+\omega_2+\omega_3=0$
and ${\bf a}=(a_1,a_2,a_3)$ describes
the strength of the fields.
Then the steady state wavefunctions $\{\psi^s ({\bf a})\}$ and action
$A^s({\bf a})$ depend also on ${\bf a}$.
Note that for ${\bf a}={\bf 0}$ the potential is time independent
and the action coincides with the static DFT energy.
By using the Hellmann-Feynman theorem we obtain the derivative of
the action with respect to the parameter $a_1$:
$$
{\partial A^s({\bf a})\over \partial a_1}=
 \int_0^T {dt\over T} \cos(\omega_1 t)\int d^3r\
{\bf e}_1\cdot {\bf r}
\ n^s({\bf r},t,{\bf a})=-{\bf e}_1 \cdot {\bf P}^s(\omega_1,{\bf a})V,
\nonumber
$$
where $V$ is the volume of the system and
${\bf P}^s(\omega_1,{\bf a})$ is the macroscopic electronic
polarization per unit volume, oscillating at frequency
$\omega_1$~\cite{footnote2}.
Then the quadratic susceptibility tensor, which is defined as
$\chi^{(2)}_{{\bf e}_1;{\bf e}_2,{\bf e}_3}(-\omega_1;\omega_2,\omega_3)
=\frac{2}{V}
\frac{\partial^2 {\bf P}^s(\omega_1,{\bf 0 })}{\partial a_2 \partial a_3}$,
is equal to:
$$
\chi^{(2)}_{{\bf e}_1;{\bf e}_2,{\bf e}_3}(-\omega_1;\omega_2,\omega_3)
=-{2\over V}
{\partial^{3} A^s({\bf 0})\over \partial a_1 \partial a_2 \partial a_3 }.
\nonumber
$$

The computation of the derivatives of $A^s({\bf a})$ with respect to $\bf a$,
can be performed by using the `$2n+1$' theorem which states that the
derivatives up to order $2n+1$ of the steady action depends only
on the change of the orbitals up to order $n$:
\begin{equation}
{\partial^{2n+1} A^s({\bf a})\over \partial {\bf a}^{2n+1} }
= {\cal P}^{2n+1}\left({\partial \{\psi^s({\bf a}) \}\over \partial {\bf a} },
\cdots,
{\partial^n \{\psi^s ({\bf a})\}\over \partial {\bf a}^n}\right),\label{2n+1}
\end{equation}
where ${\cal P}^{2n+1}$ is a polynomial of degree $2n+1$ in its arguments.
Indeed, as shown in \cite{Andrea,Langhoff}, Eq.~(\ref{2n+1}) relies just
on
the stationary condition, Eq.~(\ref{sc}).
Therefore
$\chi^{(2)}_{{\bf e}_1;{\bf e}_2,{\bf e}_3}(-\omega_1;\omega_2,\omega_3)=-
\frac{2}{V}
{\cal P}^{3}\left({\partial \{\psi^s({\bf a}) \}
\over \partial {\bf a} }\right).$
The derivation of an explicit expression of
${\cal P}^{3}$
for an infinite periodic system
requires a particular care because
the expectation value of the $\bf r$ operator
between Bloch states is ill defined.
In an insulating solid this problem can be solved following
Ref.~\cite{Andrea}:
first we apply the `$2n+1$' theorem in a Wannier representation
where the ${\bf r}$ operator is well defined,
then we recast the resulting expression in a Bloch representation.
The final expression is:
\begin{eqnarray}
\chi^{(2)}_{{\bf e}_1;{\bf e}_2,{\bf e}_3}(-\omega_1;\omega_2,\omega_3)
&=&
 -  4\sum_{m,n}^{N/2}\sum_{\sigma=\pm}
\int_{BZ} {d^3k\over (2\pi)^3}
\Bigl[
\langle u^{0}_{{\bf k},m} |
{{\bf e}_2 \over 2}\cdot{ -i\partial \over \partial {\bf k}}
\left( |u^{0}_{{\bf k},n}
\rangle \langle u^{a_1,-\sigma}_{{\bf k},n} | \right)
|u^{a_3,\sigma}_{{\bf k},m}\rangle
\nonumber\\
&&
+\delta_{m,n}\langle  u^{a_1,-\sigma}_{{\bf k},n} |V^{a_2}_{Hxc}
| u^{a_3,\sigma}_{{\bf k},m}\rangle-
\langle  u^{0}_{{\bf k},m} |V^{a_2}_{Hxc}
| u^{0}_{{\bf k},n} \rangle
\langle  u^{a_1,-\sigma}_{{\bf k},n}
|  u^{a_3,\sigma}_{{\bf k},m}\rangle
\Bigr]
\nonumber\\
&& -{4\over 6}\int d^3r d^3r' d^3r''
K_{xc}(\omega_2,\omega_3,{\bf r},{\bf r}',{\bf r}'')
n^{a_1}({\bf r})n^{a_2}({\bf r}')n^{a_3}({\bf r}'')
\nonumber\\
&&
+\Pi\{1,2,3\}.\label{chi2w}
\end{eqnarray}
Here $\Pi\{1,2,3\}$ indicates the sum over the 5 permutations of the
indexes $1,2,3$, and
\begin{eqnarray}
V^{a_1}_{Hxc}({\bf r})=\int d^3r'
\left[
{1\over |{\bf r}-{\bf r}'|}
+M_{xc}(\omega_1,{\bf r},{\bf r}')
\right]
n^{a_1}({\bf r}'),
\nonumber\\
n^{a_1}({\bf r})=2\sum^{N/2}_m
\sum_{\sigma=\pm}\int_{BZ}
{\Omega d^3k\over (2 \pi)^3}
Re\left[\langle u^{0}_{{\bf k},m}
|{\bf r} \rangle \langle {\bf r}|
 u^{a_1,\sigma}_{{\bf k},m}\rangle\right],
\nonumber\\
M_{xc}(\omega_1,{\bf r},{\bf r}') =
T\int_0^T dt {\delta^2 A_{xc}[n^{0}] \over \delta n({\bf r},0)
\delta n({\bf r}',t)}e^{i\omega_1t},
\nonumber\\
K_{xc}(\omega_2,\omega_3,{\bf r},{\bf r}',{\bf r}'')=
\int_0^T dt {\delta M_{xc}[n^0] (\omega_2,{\bf r},{\bf r}')
\over \delta n({\bf r}'',t)}e^{i\omega_3t},
\nonumber
\end{eqnarray}
$n^{0}$ is the unperturbed charge density,
$|u^{0}_{{\bf k},m} \rangle$ is the periodic part of the unperturbed
Bloch eigenstate normalized on the unit cell $\Omega$,
with eigenvalues $\epsilon^{0}_{{\bf k},m}$, and
$|u^{a_1,\pm}_{{\bf k},m}\rangle$ are the perturbed orbitals
projected on the unperturbed conduction band subspace, i.e. the solution
of the linear system:
\begin{equation}
(\epsilon^{0}_{{\bf k},m} - H^{0}_{KS}\pm \omega_1)
|u^{a_1,\pm}_{{\bf k},m}\rangle=Q_{\bf k}
\left({ {\bf e}_1\cdot {\bf r} \over 2}+V^{a_1}_{Hxc}\right)
|u^{0}_{{\bf k},m} \rangle.\label{psi1}
\end{equation}
with $Q_{\bf k}= {\bf 1}- \sum_m^{N/2}|u^0_{{\bf k},m}\rangle
\langle u^0_{{\bf k},m}|$.

Note that the evaluation of Eq.~(\ref{chi2w}) requires only the knowledge of
unperturbed valence wavefunctions $|u^{0}_{{\bf k},m}\rangle$ and of
their linear variation
$|u^{a_1,\pm}_{{\bf k},m}\rangle$.
Moreover
the solution of Eq.~(\ref{psi1}) can be obtained by minimizing a suitably
defined functional with a numerical effort
similar to the computation of the total energy ~\cite{Gonze,Pasquarello,
C60Giannozzi}.
Thus our formulation makes the evaluation of $\chi^{(2)}$
in systems containing up to an hundred atoms feasible.

We have applied Eq.~(\ref{chi2w}) to compute the static
$\chi^{(2)}$ of nine III-V and five II-VI cubic semiconductors,
evaluating the exchange and correlation energy
within the LDA.
We do not use any scissor operator to correct for the
LDA band-gap error, contrary to what has been done in other ab-initio
calculations~\cite{cinesi,Levine}.
Indeed the static $\chi^{(2)}$ is a ground state property, which is
defined as a difference of ground state total energies and it is
not related to the LDA band gap~\cite{GC}.
We think that improvements over LDA require a better
$E_{xc}$ functional,
which could be ultra-nonlocal~\cite{XG}, instead of an {\it ad-hoc}
correction of the LDA band-gap.
Furthermore our purpose here is to give reference values for the
static $\chi^{(2)}$ which are completely consistent within
LDA.

We used norm-conserving pseudopotentials and
a plane-wave kinetic energy cut-off of 24 Ry.
The derivative with respect to ${\bf k}$ which appears in Eq.~(\ref{chi2w})
has been computed by means of finite differences.
We have found that
the effect of $d$ electrons is important for Ga and
In atoms, and it necessary at least to use the nonlinear core
corrections (NLCC)~\cite{nlcc} to obtain the correct
LDA values for $\chi^{(2)}$ in the compounds containing these elements
\cite{footnoteNLCC}.
For II-VI semiconductors the effect of the cation $d$ electrons
is even more important~\cite{fononiIIVI} and our reported values
have been computed using the NLCC.
For AlP, AlAs, GaP and GaAs we have also verified that our results for the
$\chi^{(2)}$  reproduce the LDA values obtained
in Ref.~\cite{Levine} if the same pseudopotentials (without NLCC)
and lattice constants are used.

In Table~\ref{uno} we report the values of the $\chi^{(2)}$ of the III-V
and II-VI cubic semiconductors computed at the theoretical LDA
lattice constant ($a_0$), also reported in the Table.
On the same Table we show also
the direct band-gap at the $\Gamma$ point, $E_{\Gamma}$, and the
static dielectric constant $\varepsilon_{\infty}$.
Known experimental values for $a_0$, $E_{\Gamma}$ and
$\varepsilon_{\infty}$ are reported in parenthesis.
Well established experimental data for $\chi^{(2)}$ do not
exist since the values reported by different
authors may differ by more than a factor of 2.
Moreover, in some cases only data obtained at frequencies close to the
absorbtion edge are available. Therefore
we refer the readers to Ref.s~\cite{CRC,resc,cinesi}
for a complete review of the experimental results.
Just to give an indicative value,
we show in parenthesis
the experimental results from Ref.~\cite{CRC} which correspond to the lower
frequencies. For GaP, GaAs and CdSe we have taken the values
from Ref.~\cite{resc} obtained after an appropriate rescaling of the
experimental data.
In the case of InAs we cannot compute $\chi^{(2)}$ and $\varepsilon_\infty$
since within LDA the system is a metal.
For all other compounds the computed $\chi^{(2)}$ are in the range of variation
of the available experimental data \cite{CRC,resc,cinesi}.

As a second application we compute the TD
$\chi^{(2)}$ for SHG of GaP. For this calculation
we used the TD-LDA.
In the TD case the use of LDA is less justified since, in general, it
does not describe correctly the position of discrete excited levels
and absorption edges as difference to the exact TD-DFT~\cite{tddft}.
We note that this is a limitation of the approximation to $A_{xc}$ used
here, and {\it not} of Eq.~(\ref{chi2w}) itself.
Since LDA is not expected to perform sufficiently well in the TD domain
we have used the pseudopotential without NLCC which at its theoretical lattice
constant ($a_0=10.01$ a.u.) gives a gap ($E_{\Gamma}=2.8$ eV) and thus an
absorption edge,
which is incidentally close to the experimental one.

In Table~\ref{due} we report $\chi^{(2)}(2\omega;\omega,\omega)$
computed as a function of $\omega$
in the non-absorbing regime.
The experimental measurements are taken from Ref.s~\cite{resc,Levine}.

In conclusion we have presented a consistent theory for the computation
of the static and dynamic nonlinear optical susceptibilities
within DFT. To this purpose we have applied for the first time
the `$2n+1$' theorem to the TD-DFT action functional.
We have presented applications to cubic semiconductors.
Our results show that LDA reproduces the experimental
static nonlinear susceptibilities in these compounds without
using any scissor operator, provided
that the computations are performed at the theoretical
lattice constant and NLCC are included for Ga, In, Zn and Cd atoms.

We gratefully acknowledge A. Baldereschi, S. Louie,
and R. Resta for many useful discussions.
This work was supported by the Swiss National Science
Foundation under Grant No.  FN-20-30272.90, FN-21-31144.91,
and FN-21-40530.94, by the USA National  Science
Foundation under Grant No. DMR-9120269,
by DGICYT under Grant No. PB92-0645-C03-01, and by the
Miller Institute for Basic Research in Science.

\begin{table}
\caption{ LDA nonlinear susceptibilities ($\chi^{(2)}$) of III-V and II-VI
cubic semiconductors. We report also the theoretical lattice
constant ($a_0$), the direct gap at the $\Gamma$ point ($E_\Gamma$),
and the dielectric constant ($\varepsilon_\infty$). Experimental
values are given in brackets.
All computations are performed with 28 special
${\bf k}$-points, but for InSb for which we used 60 special
${\bf k}$-points. }
\begin{tabular}{lccccc}
   &$a_0$ (a.u.) & $E_\Gamma$ (eV) &$\varepsilon_\infty$
&$\chi^{(2)}$ (pm/V) \\
\tableline
AlP & 10.19 (10.33) & 3.5 (3.6) &  8.2 (7.5)  & 39  (---)   \\
AlAs& 10.56 (10.69) & 2.2 (3.1) &  9.3 (8.2)  & 64  (---)   \\
AlSb& 11.46 (11.58) & 1.9 (2.3) & 11.4 (11.3) & 146 (98)  \\
GaP&  10.12 (10.28) & 2.0 (2.9) & 10.0 (9.0)  &  83 (74)  \\
GaAs& 10.50 (10.68) & 1.0 (1.5) & 12.5 (10.9) & 205 (166) \\
GaSb& 11.37 (11.49) & 0.5 (0.8) & 16.7 (14.4) & 617 (838) \\
InP&  10.94 (11.09) & 1.0 (1.4) & 10.2 (9.6)  & 145 (287) \\
InAs& 11.34 (11.45) &-0.1 (0.4) &  --- (12.2) & --- (838) \\
InSb& 12.10 (12.23) & 0.1 (0.2) & 16.1 (15.7) & 957 (1120)\\
ZnS & 10.29 (10.22) & 2.4 (3.8) &  5.4 (5.1)  &  33 (61)  \\
ZnSe& 10.71 (10.71) & 1.6 (2.8) &  6.7 (6.3)  &  65 (156) \\
ZnTe& 11.44 (11.51) & 1.6 (2.4) &  8.1 (7.3)  & 122 (184) \\
CdSe& 11.49 (11.44) & 0.8 (1.8) &  6.9 (6.2)  & 118 (72)  \\
CdTe& 12.17 (12.24) & 1.1 (1.6) &  7.8 (7.1)  & 167 (118) \\
\end{tabular}
\label{uno}
\end{table}

\begin{table}
\caption{ The frequency dependent non linear optical susceptibility
for second harmonic generation of GaP, $\chi^{(2)}(2\omega;\omega,\omega)$.
}
\begin{tabular}{lccc}
$\chi^{(2)}$ pm/V & $\hbar \omega = 0.117$ eV  & $\hbar \omega = 0.585$ eV &
$\hbar \omega = 0.94$ eV \\
\tableline
Theo.  & 68   &   78   &  103 \\
Expt.          & 74 $\pm$ 4  &   94 $\pm$ 20 & 98 $\pm$ 18, 112 $\pm$ 12 \\
\end{tabular}
\label{due}
\end{table}


\begin{references}


\bibitem{Physicstoday} Physics Today 47, Vol.5 (1994);
J.L. Bredin, Science {\bf 263}, 487 (1994).

\bibitem{tddftperiodic} B.M. Deb and S.K. Ghosh, J. Chem Phys. {\bf 77}, 342
(1982).

\bibitem{tddft} E. Runge and E.K.U. Gross, Phys. Rev. Lett.
{\bf 52}, 997 (1984).

\bibitem{Sipe} J.E. Sipe and  Ed. Ghahramani, Phys. Rev. B {\bf 48},
11705 (1993) and references therein.

\bibitem{cinesi} M.-Z. Huang and W.Y. Ching, Phys. Rev. B {\bf 47},
9464 (1993).

\bibitem{Levine} Z.H. Levine and D.C. Allan, Phys. Rev. B {\bf 44}, 12781
(1991); Z.H. Levine, {\it ibid.} {\bf 49}, 4532 (1994); and references
therein.

\bibitem{Andrea} A. Dal Corso and F. Mauri, Phys. Rev. B {\bf 50}, 5756
(1994).

\bibitem{Langhoff} P.W. Langhoff, S.T. Epstein and M. Karplus,
Rev. Mod. Phys. {\bf 44}, 602 (1972).

\bibitem{Gonze} X. Gonze and J.P. Vigneron, Phys. Rev. B {\bf 39}, 13120 (1989)
and references therein.

\bibitem{zang} A. Zangwill and P. Soven, Phys. Rev. A {\bf 21}, 1561 (1980).

\bibitem{Sambe} H. Sambe, Phys. Rev. A {\bf 7}, 2203 (1973).

\bibitem{footnote2}
Here for simplicity we consider ${\bf P}^s(\omega_1,{\bf a})$
oscillating in phase
with the field $a_1$. This assumption is exact only in the non-absorbing
regime. To obtain $\chi^{(2)}$ in the absorbing regime
is sufficient to make the
analytical continuation of  Eq.(\protect\ref{chi2w})
to complex frequencies $\omega_1$, $\omega_2$ and $\omega_3$.


\bibitem{Pasquarello}
A. Pasquarello and A. Quattropani, Phys. Rev. B {\bf 48}, 5090 (1993).

\bibitem{C60Giannozzi} P. Giannozzi and S. Baroni, J. Chem. Phys. {\bf 100},
8537 (1994).

\bibitem{GC} A. Dal Corso, S. Baroni, and R. Resta, Phys. Rev. B
{\bf 49}, 5323 (1994).

\bibitem{XG} X. Gonze, Ph. Ghosez, and R.W. Godby, Phys. Rev. Lett. {\bf 74},
4035 (1995).

\bibitem{Bachelet} G.B. Bachelet, D.R. Hamann, and M. Sch\"ulter,
Phys. Rev. B {\bf 26}, 4199 (1982).

\bibitem{CRC} S. Singh, in {\it Handbook of Laser Science and Technology},
Ed. M.J. Weber, Boca Raton, FL: CRC, 1986, vol III.

\bibitem{resc} D.A. Roberts, IEEE Jour. of Quan. Electr. {\bf 28}, 2057 (1992).

\bibitem{nlcc} S.G. Louie, S. Froyen, and M.L. Cohen, Phys. Rev. B
{\bf 26}, 1738 (1982).

\bibitem{footnoteNLCC} The $\chi^{(2)}$ of InSb computed neglecting the NLCC
is, e.g., one half of the LDA value reported in Tab.~\protect\ref{uno}.

\bibitem{fononiIIVI} A. Dal Corso, S Baroni, R. Resta and  S. De Gironcoli,
Phys. Rev. B {\bf 47}, 3588 (1993).

\end{references}
\end{document}